\documentclass[fleqn,twoside,twocolumn,nofootinbib,showkeys]{revtex4} % Specifies the document class %,unsortedaddress
\usepackage[sec,nocpr]{ujp} % \usepackage[cyr]{ujp} for cyrillic, \usepackage[web]{ujp} for web
\begin{document}
\title [Energy
Terms and Stability Diagrams] {ENERGY TERMS AND STABILITY\\ DIAGRAMS
FOR THE 2\boldmath$D$ PROBLEM\\ OF THREE CHARGED PARTICLES}
\author{I.V.~Simenog}
\affiliation{\bitp}
\address{\bitpaddr}
\email{ivsimenog@bitp.kiev.ua}
\author{V.V.~Mikhnyuk}
\affiliation{\bitp}
\address{\bitpaddr}
\email{ivsimenog@bitp.kiev.ua}
\author{Yu.M.~Bidasyuk}
\affiliation{\bitp}
\address{\bitpaddr}
\email{ivsimenog@bitp.kiev.ua}

\udk{???}  \razd{\secix}

\autorcol{I.V.\hspace*{0.7mm}Simenog, V.V.\hspace*{0.7mm}Mikhnyuk,
Yu.M.\hspace*{0.7mm}Bidasyuk}

%\makeatletter
%\renewcommand{\thesection}{\arabic{section}}
%\renewcommand{\p@subsection}{}
%\renewcommand{\thesubsection}{\arabic{section}.\arabic{subsection}}
%\renewcommand{\p@subsubsection}{}
%\renewcommand{\thesubsubsection}
%{\arabic{section}.\arabic{subsection}.\arabic{subsubsection}}
%\makeatother

%\input{tcilatex}

\setcounter{page}{439}%

\begin{abstract}
Symmetric and antisymmetric terms have been obtained in the
framework of the variational approach for two-dimensional (2$D$)
Coulomb systems of symmetric trions $XXY$.\,\,Stability diagrams and
certain anomalies arising in the 2$D$ space are explained
qualitatively in the framework of the Born--Oppenheimer adiabatic
approximation.\,\,The asymptotics of energy terms at large distances
obtained for an arbitrary space dimensionality are analyzed, and
some approximation formulas for 2$D$ terms are proposed.\,\,An
anomalous dependence of multipole moments on the space
dimensionality has been found in the case of a spherically symmetric
field.\,\,The main results obtained for the 2$D$ and 3$D$ problems
of two Coulomb centers are compared.
\end{abstract}


\maketitle

\section{Introduction}

The two-dimensional (2$D$) problems for Coulomb systems arise in
various physical domains: for layered and near-surface materials, in
the physics of graphene, and in connection with general problems
aimed at studying the dependences of physical observables on the
space dimensionality (e.g., see works \cite{R1,R2}).\,\,Researches
of conditions required for the emergence of bound states in the 2$D$
problem of three charged particles and their comparison with those
for the same problem in the three-dimensional (3$D$) space
\cite{R4,R5} revealed certain important anomalies in the 2$D$ space,
especially in the molecular mode \cite{R3}.\,\,Hence, there arises a
necessity of a clearer, at the physical level, understanding of such
two-dimensional features that are absent in the standard formulation
of corresponding problems in the 3$D$ space.\,\,In the molecular
mode for a symmetric trion $XXY$ (in this case, we actually deal
with two Coulomb centers), this 3$D$ problem has been studied rather
completely (in particular, see works \cite{R61,R62,R6}).\,\,For
today, separate results for the asymptotics of energy terms at short
and large distances have already been obtained for two Coulomb
centers in the 2$D$ space as well \cite{R1,R2}.\,\,However, the
application of these terms to studying the stability diagrams still
requires a separate consideration.

In this work, the energy terms were calculated in the framework of a
rather accurate variational approach and without separating the
variables.\,\,This technique can be extended to include more
complicated systems.\,\,Some approximation formulas for the terms
are proposed with regard for the asymptotic formulas for large and
short distances between the centers, and the features of the terms
specific for various space dimensionalities are discussed.\,\,The
obtained terms are used to analyze the stability diagrams plotted in
the mass--charge, $(m,Z),$ plane in the Born--Oppenheimer adiabatic
approximation.\vspace*{-2mm}

\section{Variational Calculation of Energy Terms}

The Hamiltonian of the 2$D$ symmetric problem of $XXY$ for three
charged particles looks like\vspace*{-2mm}
%1
\begin{equation} \label{eq:3pH}
 \hat{H}=\dfrac{\hat{p}_1^2+ \hat{p}_2^2}{2m}+\dfrac{\hat{p}_3^2}{2}+
\dfrac{1}{r_{12}}-Z\left(\! \dfrac{1}{r_{13}}+
 \dfrac{1}{r_{23}}\!\right)\!,
 \end{equation}
where the standard expression
%2
\begin{equation}
V_{C}=\frac{1}{r}  \label{eq:V_C}
\end{equation}%
is taken for the Coulomb potential in isolated small systems with a
plane geometry.\,\,Concerning the rather delicate issues dealing
with the choice of an interaction potential between charged
particles in low-dimensional systems, we confine the discussion to
the remark that this problem in the case of thin films was
considered long ago in work \cite{R7}, in work \cite{R8} at a more
simplified level, and in recent work \cite{R81}.\,\,The application
of potential (\ref{eq:V_C}) to small systems of charged particles
located on a plane (2$D$ problem) can be justified by the fact that
the system looks as if it is confined within a thin layer,
i.e.\,\,the system motion in one of the directions is restricted by
a very narrow and deep quantum well.\,\,The thickness of this layer
does not appear in subsequent calculations, but its allowable values
can be estimated from the form of energy terms.\,\,A finite
thickness deforms the energy terms in the 2$D$ problem only at
distances comparable with this thickness, and the repulsive region
at short distances does not substantially affect the vibration
spectrum of the term.\,\,Therefore, the layer thickness must be
narrower than the repulsive region of the term (see the distance
$R_{0}$ in Table~\ref{tab:saparameters}).\,\,Then, the energy
spectrum of the system will not differ substantially in this case
from that obtained in the 2$D$ problem.\,\,Hence, we assume the
motion along the direction $y$, i.e.\,\,perpendicularly to the
plane, to be frozen, and the distance in the plane to be determined
by the formula
%3
\begin{equation}
r^{2}=x^{2}+z^{2}.
\end{equation}

Let us rewrite Eq.~(\ref{eq:3pH}) in the center-of-mass system, i.e. in
the relative coordinates
%4
 \begin{equation}
\begin{array}{l}
 \displaystyle\mathbf{R}=(\mathbf{r_1}-\mathbf{r_2})Z\!\left(\!1+\frac1{2m}\!\right)^{\!\!-1}\!\!,
 \\[3mm]
\displaystyle\mathbf{r}=\left(\mathbf{r_3}-\frac{\mathbf{r_1}+\mathbf{r_2}}
{2}\right)\!Z\!\left(\!1+\frac1{2m}\!\right)^{\!\!-1}\!\!,
\end{array}
\end{equation}
where ${\mathbf{R}}$ is the radius-vector describing the relative
position of one identical particle with respect to the other
one.\,\,Then, the Schr\"{o}dinger equation looks like
%5
\[
\biggl\{\!
-\dfrac{1}{m+1/2}\Delta_R-\dfrac{1}{2}\Delta_r+\dfrac{1}{ZR}-\dfrac{1}
{|\mathbf{r}-\mathbf{R}/2|} \,-
\]%\vspace*{-7mm}
\begin{equation}
 \label{eq:BO}
-\,\dfrac{1}{|\mathbf{r}+\mathbf{R}/2|}\! \biggr\}
\Psi(r,R)=\boldsymbol{\epsilon}  \Psi(r,R),
\end{equation}
which is a convenient form for the separate analysis of the fast
electron motion described by the vector ${\bf r}$ and the slow (in
the molecular mode, when \mbox{$m\gg 1$}) vibration motion described
by the vector ${R}$.\,\,The energy $E$ in problem (\ref{eq:3pH}) is
determined through the energy of Eq.~(\ref{eq:BO}) as follows:
%6
\begin{equation}
E=\dfrac{2mZ^{2}}{1+2m}\,\boldsymbol{\epsilon }.
\end{equation}
Consider two heavy particles in the molecular mode, when the
Born--Oppenheimer (BO) adiabatic approximation, which allows the
fast electron and slow vibration motions to be analyzed separately,
is applicable, so that
%7
 \begin{equation}
 \Psi(\mathbf{r},\mathbf{R})\approx \Phi(\mathbf{r},\mathbf{R})\chi(\mathbf{R}),
 \end{equation}
where $\Phi ({\mathbf{r}},{\mathbf{R}})$ is the electron wave
function of the fast coordinate ${\mathbf{r}}$ at fixed
${\mathbf{R}}.$\,\,Then the Schr\"{o}dinger equation for the
electron dynamics is a two-dimensional two-center Coulomb eigenvalue
problem,
%8
 \begin{equation} \label{eq:e}
\left\{\!-\dfrac{1}{2}\Delta_r-\dfrac{1}{|\mathbf{r}|}-
\dfrac{1}{|\mathbf{r}+\mathbf{R}|}\!\right\}\!\Phi(\mathbf{r},\mathbf{R})=U(R)\Phi(\mathbf{r},\mathbf{
R}),\!\!\!\!
\end{equation}
which must be solved to determine the terms $U(R)$.\,\,In addition,
the terms $U_{(s,a)}(R)$ must additionally correspond to the
symmetric, $s$, or antisymmetric, $a$, states with respect to
permutations of identical Coulomb centers.\,\,Note that the
two-center problem (\ref{eq:e}) can be considered in the space of
any dimensionality, which is of interest for the analysis of
specific features revealed by the solutions and depending on the
space dimensionality, and this sheds light on anomalies arising in
the 2$D$ problem.\,\,While determining the energy states in the BO
approximation, the terms used as the effective interaction
potentials in the Schr\"{o}dinger equation for vibration spectra are
as follows:
%9
\[
\left\{\!-\dfrac{1}{m+1/2}\Delta_R+\dfrac{1/Z-1}{R}+V_{(s,a)}(R)\!\right\}\chi_{n(s,a)}(R)=
\]\vspace*{-7mm}
\begin{equation} \label{eq:eq9}
=\varepsilon\chi_{n(s,a)}(R),
\end{equation}
where\vspace*{-2mm}
%10
\begin{equation} \label{eq:saterms}
V_{(s,a)}(R)=U_{(s,a)}(R)-U_{(s,a)}(\infty)+\frac1R.
\end{equation}
In order to find the terms $U(R)$ in the two-center problem
(\ref{eq:e}), the separation of variables in the ellipsoid
coordinates is conventionally considered (see, e.g., work \cite{R6}
for the 3$D$ problem and work \cite{R2} for the 2$D$ one), and the
solutions of a system of two one-dimensional equations are
analyzed.\,\,In this work, in order to find the solutions of the
two-center 2$D$ problem (\ref{eq:e}), we use an alternative
variational method.\,\,We hope to extend this approach in the future
to relativistic problems, problems with a larger number of centers,
and systems with more than three particles, for which the separation
of variables is impossible.\,\,To determine the eigenvalues of
problem (\ref{eq:e}) for various center-to-center distances $R$, we
use the Galerkin variational method with the basis functions (with
the violated spherical or polar symmetry, of course)
%11
\begin{equation} \label{eq:totalfunction}
\phi_i=e^{-a_ix^2}\left(\!e^{-b_iz^2}+se^{-b_i\left(z+R\right)^2}\!\right)\!,
\end{equation}
where $s=+1$ for symmetric and $s=-1$ for antisymmetric states with
respect to the permutation of centers.\,\,In addition, to make the
consideration more general, we will analyze the $d$-dimensional
problem with an arbitrary space dimension $d$, although the specific
calculations will be carried out in the 2$D$ case.\,\,Then
%12
\begin{equation}
x^{2}=\sum\limits_{i=1}^{d-1}x_{i}^{2}.
\end{equation}
The total variational function is taken in the form
%13
\begin{equation}
\Phi =\sum\limits_{i=1}^{K}N_{i}\phi _{i},
\end{equation}%
and the corresponding spectra for the terms in the two-center problem
(\ref{eq:e}) and the eigenfunctions are determined by solving the system of linear
algebraic equations
%14
\begin{equation} \label{eq:linear_eqs}
\sum\limits_{j=1}^{K}N_j\left\{\! \langle \phi_i\left|
 \hat{H}-E\right|\phi_j \rangle\!\right\}=0,\quad i=\overline{1{,}K}.
\end{equation}
Note that, to make further calculations more convenient, the
coordinate origin in Eqs.~(\ref{eq:e}) and (\ref{eq:totalfunction})
is placed at one of the centers rather than at the middle point
between them, the latter seems to be natural.\,\,Then, the general
energy matrix in Eq.~(\ref{eq:linear_eqs}) for the $d$-dimensional
problem calculated with the use of the basis functions
(\ref{eq:totalfunction}) reads
%\begin{widetext}
%15
\[
\left<\phi_i|H-E|\phi_j\right>=(d-1)\dfrac{a^ia^j}{a^{(d+1)/2}b^{1/2}}\,\times
\]\vspace*{-7mm}
\[
 \times\left(\!1+s \exp{\left\{\!-\dfrac{b^ib^j}{b}R^2
\!\right\}}\!\right)+ \dfrac{b^ib^j}{a^{(d-1)/2}b^{3/2}}\,\times
\]\vspace*{-7mm}
\[
\times\left(\!1+s\left(1-\dfrac{2b^ib^j}{b}R^2\!\right)\exp{\left\{\!-\dfrac{b^ib^j}{b}R^2\!\right\}}\!\right)-
\]\vspace*{-7mm}
\begin{equation}
-I-E\dfrac{\left(\!1+s\exp{\left\{\!-\dfrac{b^ib^j}{b}R^2\!\right\}}\!\right)}{a^{(d-1)/2}b^{1/2}},
\end{equation}
where the integral $I$ for the potential energy of two centers is determined
as follows:
%16
 \[
 I=\dfrac{4}{\sqrt{\pi}}\int\limits_{0}^{1}dx\dfrac{x^{d-2}(2-x^2)^{(d-3)/2}}{\sqrt{(a+(b-a)(1-x^2)^2)}}\,
\times
\]\vspace*{-7mm}
\[
\times\biggl\{1+\exp{\left\{-b(1-x^2)^2R^2\right\}}+s\,\times
\]\vspace*{-7mm}
\[
\times\biggl(\!\exp{\left\{\!-\dfrac{b^j(b^i+b^j(1-x^2)^2)}{b}R^2\!\right\}}+
\]\vspace*{-7mm}
\begin{equation} \label{eq:integrals}
+\exp{\left\{\!-\dfrac{b^i(b^j+b^i(1-x^2)^2)}{b}R^2\!\right\}}\!\biggr)\!\biggr\}.
\end{equation}

%рис.1
\begin{figure} [t]
\vskip1mm
 \includegraphics[width=\column]{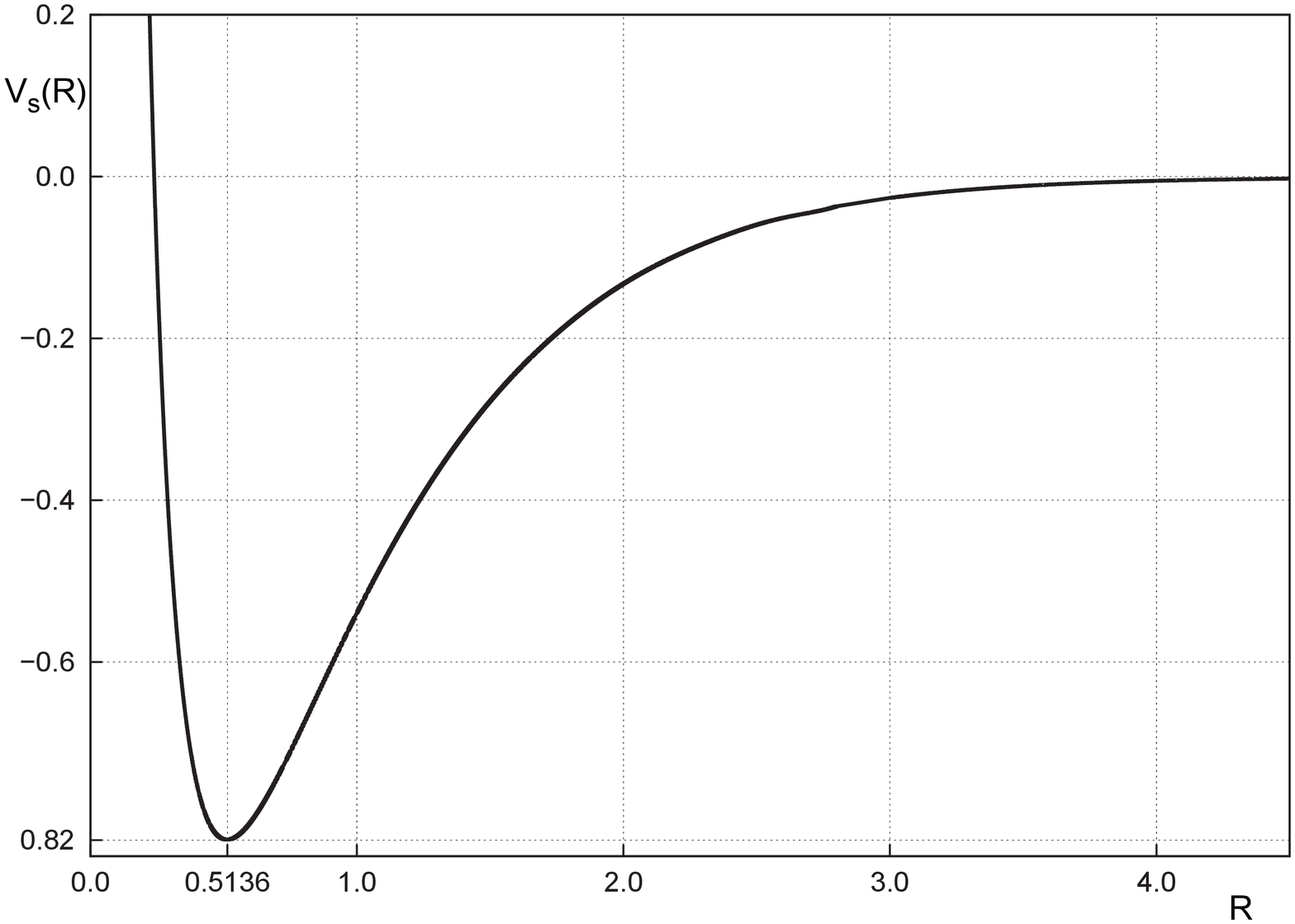}
 \vskip-3mm\caption{Symmetric electron term
($V_{s}(R)=U_{s}(R)+2.0+1/R$)} \label{fig:sterm}
\end{figure}

%рис.2
\begin{figure} [t]
\vskip1mm
\includegraphics[width=\column]{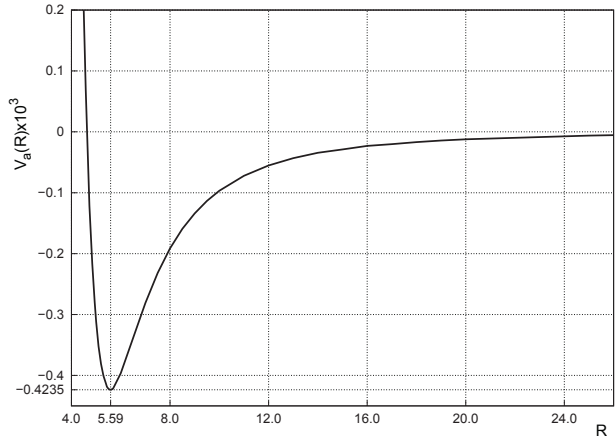}
\vskip-3mm\caption{Antisymmetric electron term
($V_{a}(R)=U_{a}(R)+2.0\,+$ $+\,1/R$)}
\label{fig:aterm}\vspace*{-2mm}
\end{figure}

%Табл. 1

\begin{table}[b]
\noindent\caption{Parameters of symmetric\\ (\boldmath$s$) and
antisymmetric ($a$) terms in 2$D$\\ and 3$D$
problems}\vskip3mm\tabcolsep9.2pt

\noindent{\footnotesize\begin{tabular}{|c|c|c|c|c|}
 \hline \multicolumn{5}{|c|}
{\rule{0pt}{4mm}$s$} \\[1.5mm]%
\hline%
 \multicolumn{1}{|c|}{}&\multicolumn{1}{|c|}{\rule{0pt}{4mm}$R_{\min}$}&
\multicolumn{1}{|c|}{$V_{\min}$}& \multicolumn{1}{|c|}{$R_{0}$}&
\multicolumn{1}{|c|}{$R_{13}$}\\[1.5mm]%
\hline%
\rule{0pt}{4mm}$2D$&0.51357&$-$0.820~\,~\,~\,&0.2391&0.5821\\%
$3D$&1.99719&$-$0.102635&1.10~\,~\,&\\[1.5mm]%
\hline%
 \multicolumn{5}{|c|}{\rule{0pt}{4mm}$a$}\\[1.5mm]%
\hline%
\multicolumn{1}{|c|}{}&\multicolumn{1}{|c|}{\rule{0pt}{4mm}$R_{\min}$}&
\multicolumn{1}{|c|}{$V_{\min}$}& \multicolumn{1}{|c|}{$R_{0}$}&
\multicolumn{1}{|c|}{$R_{13}$}\\[1.5mm]%
\hline%
\rule{0pt}{4mm}$2D$&~\,5.59&  $-4.235\times10^{-4}$&~\.4.625&3.9994\\%
$3D$&12.55& ~\,$-6.08\times10^{-5}$&10.69~\,&\\[2mm]%
\hline
\end{tabular}}
\label{tab:saparameters}
\end{table}

\noindent The features in the solution of problem
(\ref{eq:linear_eqs}) in the BO adiabatic approximation are
analogous to the difficulties faced with when solving the problem of
three particles, irrespective of the difficulties in the numerical
calculation of integrals (\ref{eq:integrals}); however, we will not
discuss this issue.

The results of calculations carried out for the lowest terms (the
ground states) in the symmetric and antisymmetric states, which were
obtained for the 2$D$ problem, are depicted in Figs.~\ref{fig:sterm}
and \ref{fig:aterm}, respectively.\,\,The results of calculation for
the symmetric term completely agree with the result of work
\cite{R9} obtained in a different way.\,\,Note that, in this
formulation, the polar (spherical) symmetry is absent, and, from the
general point of view, all states can be classed in the
hyperspheroidal coordinates, where the total separation of variables
can be executed (see works \cite{R9,R1,R2}).\,\,For convenience, we
presented terms (\ref{eq:saterms}) in such a form that they vanish
as $R\rightarrow \infty $, which is convenient, while considering
effective interaction potentials.\,\,The value $U_{(s,a)}(\infty
)=-2.0$ is the energy of one center (the hydrogen atom) in the
ground state.\,\,This is a two-particle decay threshold.\,\,We would
like to emphasize that the both terms $V_{(s,a)}(R)$ calculated for
the lowest states in the 2$D$ problem remain negative (attraction)
at significant distances ($R\gg 1$), as well as doubly degenerate in
this limit (we consider the two-center problem), the same being also
valid for the 3$D$ space.\,\,At short distances (the two centers are
united, $U_{(s)}(0)=-8,$ and $U_{(a)}(0)=-8/9$), the terms
$V_{(s,a)}(R\rightarrow 0)$ are positive (repulsion) owing to the
repulsive Coulomb potential between the identical centers.

The characteristic parameters of symmetric and antisymmetric 2$D$
terms (attractive potential wells) are quoted in
Table~\ref{tab:saparameters}, where a comparison with the
corresponding parameters for the 3$D$ problem~\cite{R5} is also
made.\,\,The Table demonstrates the positions of the minima,
$R_{\mathrm{min}}$, the values of terms at the corresponding minima,
$V_{\min }$, and the distance $R$, above which the terms are
negative.\,\,We would like to emphasize that the symmetric 2$D$ term
is almost eight times as deep as the 3$D$ term (stronger coupling),
and its position is located approximately four times nearer.\,\,The
distance $R_{0}$ in the 2$D$ case is also approximately four times
shorter than that in the 3$D$ problem.\,\,The minimum value of
antisymmetric 2$D$ term is anomalously large in comparison with the
3$D$ term, but is much smaller than the value of symmetric term.
This means that, similarly to the 3$D$ space, the conditions for the
emergence of antisymmetric states are much poorer than the
conditions, at which the symmetric states with the given charge,
$Z$, and mass, $m$, values exist.\,\,However, in the 2$D$ space, the
antisymmetric term corresponds to a much stronger attraction than in
the 3$D$ state.

Note that, in the BO approximation, the root-mean-square distance
between the third (light) particle and one of the centers, $R_{31}$,
which can be calculated using the electron functions $\Phi
({\mathbf{r}},{\mathbf{R}}_{\mathrm{min}})$ according to the formula
%17
 \begin{equation}
 R^2_{31}\equiv\langle R^2_{31}\rangle={\int d\mathbf{r}R^2_{31}\Phi^2(\mathbf{r},
 \mathbf{R}_{\min})}/{\int d\mathbf{r}\Phi^2(\mathbf{r},\mathbf{R}_{\min})}
 \end{equation}
satisfies the abnormal relation
%18
\begin{equation}
R_{31}>R_{\mathrm{min}}
\end{equation}%
at the point $R_{\mathrm{min}}$ in the case of a symmetric main
term. This relation was revealed in work \cite{R3}, while carrying
out three-particle calculations.\,\,Hence, the fact is confirmed
that the distance between the light particle and the attracting
center calculated in the 2$D$ problem for large masses of two
centers in molecular systems exceeds the distance between repulsive
centers (let it be determined as $R_{\mathrm{min}}$).\,\,However,
the natural relation $R_{31}<R_{\mathrm{min}}$ takes place in the
case of the 3$D$ problem.\,\,In the 2$D$ problem, the natural
relation is also obeyed for the antisymmetric state both in
three-particle calculations \cite{R3} and in the BO approximation.

\begin{figure}[t]
\vskip1mm
\includegraphics[width=\column]{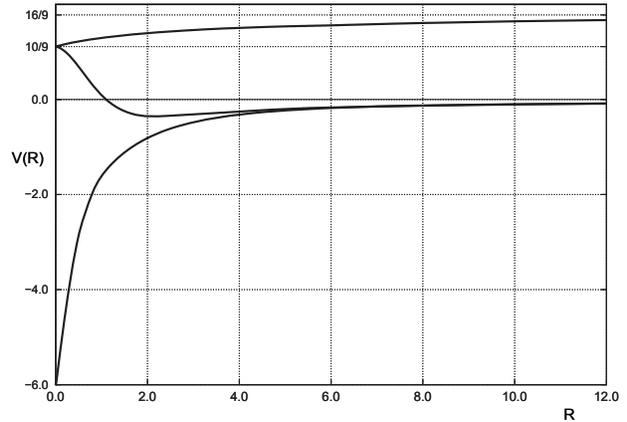}
\vskip-4mm\caption{Ground ($s_{0},a$) and excited ($s_{1}$) electron
terms } \label{fig:saterms}\vspace*{-1mm}
\end{figure}

It is worth to note that, if considering the vibration states in the
BO approximation, it is reasonable to consider only the terms
indicated for the symmetric and antisymmetric ground states, because
all other excited terms lie above the two-particle decay threshold
$E_{0}(2)=-2.0$.\,\,In Fig.~\ref{fig:saterms}, the main terms
$s_{0}$ and $a_{0}$, as well as the excited symmetric term $s_{1}$
(designated as $U(R)$ for convenience), are shifted by the
two-particle threshold energy, but are not shifted by the Coulomb
potential, analogously to the 3$D$ problem.\,\,The figure makes it
evident that only the main terms can be responsible for the
appearance of bound states in Eq.~(\ref{eq:eq9}).

\section{Term Asymptotics}

Consider the asymptotics of 2$D$ terms at large and short distances,
which are obtained according to the Schr\"{o}dinger equation
(\ref{eq:e}), and, if possible, let us make a generalization onto an
arbitrary space dimensionality $d$.\,\,This problem was already
examined in a series of works by Lazur and coworkers \cite{R1,R2}.
In this work, we will pay more attention to physical conclusions.
The $d$-dimensional two-center Coulomb problem allows the separation
of all variables in the hyperspheroidal coordinates; these are two
linear coordinates, $\xi =(r_{1}+r_{2})/R$ and $\eta
=(r_{1}-r_{2})/R$, and $d-2$ angular variables, in a full analogy
with the 3$D$ space (see, e.g., works \cite{R5,R9}).\,\,In the space
of $\xi $ and $\eta $ coordinates, the problem for the lowest states
that are independent of the angles is reduced to the following
system of two coupled one-dimensional equations:
%19
\begin{equation} \label{eq:systemseq}
\begin{array}{l}
 \displaystyle
(\xi^2-1)\dfrac{d^2X}{d\xi^2}+2\sigma\xi
 \dfrac{dX}{d\xi}\,+\\[3mm]
\displaystyle +\left(\dfrac{ER^2}{2}(\xi^2-1)+2R(\xi-1)+2R+RA\right)X=0,\\[3mm]
\displaystyle (1-\eta^2)\dfrac{d^2Y}{d\eta^2}-2\sigma\eta\dfrac{dY}{d\eta}\,+\\[3mm]
\displaystyle +\left(\dfrac{ER^2}{2}(1-\eta^2)-RA\right)Y=0,
 \end{array}\!\!\!\!\!\!\!\!\!\!\!\!\!\!\!\!\!\!\!\!\!\!
\end{equation}
where $\xi \geq 1$, $-R<\eta <R$, $U(R)$ is the term, $A$ is the separation
parameter, and $\sigma =(d-1)/2$ defines the space dimensionality.

%рис.3

In all aspects, the system of one-dimensional equations
(\ref{eq:systemseq}) is similar to the known problem in the 3$D$
space.\,\,Therefore, to derive an analytical solution, a standard
perturbation theory can be developed for large and small distances
$R$.\,\,At large distances, where the symmetric and antisymmetric
terms are degenerate to within an exponential accuracy, it is
convenient to introduce the variables $\xi =1+\dfrac{2x}{R}$ ($x\geq
0$), and $\eta =1-\dfrac{2y}{R}$ ($0\leq y\leq R$) and, changing the
variables,
%20
\begin{equation}
% \begin{split}
X=\exp{\Bigg\{\!\!-\!\int \limits^x\! C(x')dx'\!\Bigg\}},~
Y=\exp{\Bigg\{\!\!-\!\int\limits ^y\! D(y')dy'\!\Bigg\}},
% \end{split}
\end{equation}
to consider the equations of the Riccati type,
%21
\begin{equation} \label{eq:rikkati}
\begin{array}{l}
\displaystyle
x\left(1+\dfrac{x}{R}\right)\left(-\dfrac{dC}{dx}+C^2\right)-\sigma\left(1+\dfrac{2x}{R}\right)\times\\[3mm]
\displaystyle \times\, C+2+A+2Ex\left(1+\dfrac{x}{R}\right)+\dfrac{4x}{R}=0,\\[3mm]
\displaystyle
y\left(1-\dfrac{y}{R}\right)\left(-\dfrac{dD}{dy}+D^2\right)-
\sigma\left(1-\dfrac{2y}{R}\right)\times\\[3mm]
 \displaystyle \times\, D-
A+2Ey\left(1-\dfrac{y}{R}\right)=0.
 \end{array}
\end{equation}
We will use the so-called logarithmic perturbation theory in the
small parameter $1/R$, when the energy, separation parameter, and
solutions are sought in the form of a power series in the small
parameter $1/R$ and satisfy equations of the Riccati type (see
Eqs.~(\ref{eq:rikkati})).\,\,The final power series for the
symmetric term of the ground state (as well as for the antisymmetric
term, which is degenerate with the former in the power-law
approximation) obtained from corresponding recurrent relations in
the space of any dimensionality $d>1$ looks like
%22
\[
U_0(R)=-\frac1{2\sigma^2} -\frac1R
-\frac1{R^3}\frac12\sigma^2(1-\sigma^2)\,-
\]\vspace*{-7mm}
\[
-\,\frac1{R^4}\frac18\sigma^4(1+\sigma)(5+4\sigma)-\frac1{R^5}
\frac38 \sigma^4(1-\sigma^2)(4-\sigma^2)\,-
\]\vspace*{-7mm}
\begin{equation}
\label{eq:u0}
-\,\frac1{R^6}\frac14\sigma^6(1+\sigma)(28+5\sigma-14\sigma^2-4\sigma^3)+O\left(\!\frac1{R^7}\!\right)\!,
\end{equation}
where four first terms reproduce the terms obtained in work
\cite{R1}.\,\,The logarithmic derivatives of wave functions,
obtained in the framework of perturbation theory, are
%23
\[
C(x,R)=\frac1{\sigma}+\frac1{R^2}\sigma\left(\!x+\frac{\sigma}2(1+\sigma)\!\right)
-\frac1{R^3}\sigma x(x+\sigma)\,+
\]\vspace*{-7mm}
\[
+\,\frac1{R^4}\sigma\biggl[x^3+x^2\frac{\sigma}2(3-2\sigma)-x\frac{\sigma^2}4\biggl(\!6
\sigma^2\,+
\]\vspace*{-7mm}
\begin{equation}
+\,7\sigma-6\!\biggr)+\frac38\sigma^3(1-\sigma^2)(2+\sigma)\biggr]+O\left(\!\frac1{R^5}\!\right)\!,
\end{equation}\vspace*{-7mm}
%24
\[
D(x,R)=\frac1{\sigma}-\frac1{R^2}\sigma\left(\!y+\frac{\sigma}2(1+\sigma)\!\right)
-\frac1{R^3}\sigma y(y+\sigma)\,-
\]\vspace*{-7mm}
\[
-\,\frac1{R^4}\sigma\biggl[y^3+y^2\sigma\frac32+y\frac{\sigma^2}4(6+\sigma)\,+
\]\vspace*{-7mm}
\begin{equation}
+\,\frac38\sigma^3(1-\sigma^2)(2+\sigma)\biggr]+O\left(\!\frac1{R^5}\!\right)\!.
\end{equation}
However, the shift of the antisymmetric term with respect to the symmetric one
has an exponential character irrespective of the space dimensionality,
%25
\begin{equation} \label{eq:deltaE}
\delta E =\frac{16}{\sigma^3
\Gamma(\sigma)}\left(\!\frac{R}{2\sigma}\!\right)^{\!\!\sigma}
e^{-R/\sigma-\sigma}\left[1+\frac{\sigma}{2R}+O\left(\!\frac1{R^2}
\!\right)\!\right],
\end{equation}
in accordance with the results of works \cite{R1,R5,R8} (here,
$\Gamma (\sigma )$ is the gamma-function).

We would like to make a few general remarks.\,\,First, the 2$D$
terms (\ref{eq:saterms}), taking into account Eq.~(\ref{eq:u0}),
have an abnormal attractive asymptotics (\ref{eq:u0}) of an order of
$-1/R^{3}$.\,\,Second, the asymptotic series (\ref{eq:u0}) and, the
more so, the power series near the exponent in Eq.~(\ref{eq:deltaE})
always diverge following the factorial law, which is well known to
be a general rule for a diversity of problems in the 3$D$ space.
Hence, such series can be used for estimations only at large enough
distances.\,\,Third, from the formal viewpoint, the computer
facilities allow a significant number of higher power terms in
series (\ref{eq:u0})--(\ref{eq:deltaE}) to be calculated, but,
generally speaking, this procedure has no reason because of the
series divergence.\,\,By the way, the character of divergence for
such asymptotic series becomes weakened a little as the space
dimensionality $d$ diminishes.\,\,In particular, in the limit
$d\rightarrow 1$, provided that $d$ is strictly larger than 1 and
that this case has a physical sense, only the first terms in
Eq.~(\ref{eq:u0}) would expectedly compose a good approximation for
the term, even at moderate values of the distance $R$.

We would like to attract attention to one more important conclusion.
In the 2$D$ space, the main asymptotics at large distances for the
interaction potential of the van der Waals type between neutral
atoms in ground states with zero angular momenta is the abnormal
repulsion law $C/R^{5}$ rather than the standard attraction one
$-A/R^{6}$ (in the 3$D$ space) even in the first order of
perturbation theory.\,\,In particular, in the 2$D$ space, as well as
for an arbitrary dimensionality $d$, we have a repulsive potential
of interaction of the quadrupole-quadrupole and quadrupole-dipole
types between two hydrogen atoms at very large distances,
\[
\dfrac{1}{R^5}\biggl[\dfrac{3}{4}\langle \hat{Q}_2(1) \rangle
\langle\hat{Q}_2(2)\rangle +\dfrac{1}{2}\biggl(\!\!\langle
\hat{d}(1)^2\rangle \langle\hat{Q}_2(2)\rangle\,+
\]\vspace*{-7mm}
\[
+\,\langle\hat{d}(2)^2\rangle\ \langle\hat{Q}_2(1)\rangle
\!\biggr)\!\biggr]=\dfrac{123}{1024}\dfrac{1}{R^5}.
\]
Moreover, such abnormal repulsive asymptotics is of high importance
in a space with a large enough dimensionality $d$, rather than in
the 3$D$ one, where it vanishes, and in spaces with $d$ slightly
exceeding 3, where a weak attractive asymptotics $-C/R^{5}$ is
realized.\,\,It should be recalled that, in this case, there also
exists a centrifugal barrier of the kinematic origin,
$(d-3)(d-1)/4R^{2}$.\,\,The next asymptotic term, $-A/R^{6}$, which
is obtained in the second-order perturbation theory in the
dipole-dipole interaction parameter, is always attractive for an
arbitrary space dimensionality.\,\,Moreover, the constant $A$, most
likely, can be much larger that the constant $C$ (in the term
$-C/R^{5}$).\,\,Therefore, the manifestation of the repulsive
asymptotics in bound states can be strongly suppressed at a low
dimensionality.\,\,This conclusion is also supported by the presence
of the centripetal attraction of the kinematic origin.

A separate attention should be attracted to the fact that the terms
with the third, fifth, and some higher odd power exponents of the
reciprocal radius in Eq.~(\ref{eq:u0}) are contributions of the
first-order perturbation theory.\,\,They are determined by the
contributions of a quadrupole (the $1/R^{3}$-term in
Eq.~(\ref{eq:u0}), it equals zero only in the 3$D$ space), octupole
(the $1/R^{5}$-term), and sixth-order multipole (some part of the
$1/R^{7}$-term) averaged over the wave function of one-center
problem.\,\,The $1/R^{4}$-term is a contribution of the second-order
perturbation theory and describes the dipole-dipole
interaction.\,\,The $1/R^{6}$-term is a superposition of the
dipole-octupole and quadrupole-quadrupole interactions in the
second-order perturbation theory.\,\,The remaining part of the
$1/R^{7}$-term arises owing to the contribution of the
dipole-quadrupole interaction in the third-order perturbation
theory, and so on.\,\,Really, the following expansion into a
multipole series is valid at large distances $R$ between the
centers:
%26
\[
 \frac1{|\mathbf r- \mathbf
R|}=\frac1R\sum\left(\!\frac{r}{R}\!\right)^{\!k}P_k
\left(\!\frac{(\mathbf r \mathbf R)}{rR}\!\right)=
\]\vspace*{-5mm}
\begin{equation} \label{eq:1/|r-R|}
=\frac1R\left(\!1+\frac{\hat{d}}{R}+\frac{\hat{Q}_2}{2R^2}+\sum
\frac{\hat{Q}_k}{R^k}\!\right)\!,
\end{equation}
where $P_{k}$ are the Legendre polynomials depending on the angle
cosine $\cos (\phi
)=({\mathbf{r}}{\mathbf{R}})/|{\mathbf{r}}||{\mathbf{R}}|$,
$\hat{d}=z=r\cos (\phi )$ is the dipole moment operator,
$\hat{Q}_{2}=(3z^{2}-r^{2})$ is the quadrupole moment operator, and
$\hat{Q}_{k}=r^{k}P_{k}(\cos (\phi ))$ are the operators of
higher-order multipole moments.\,\,Then the even multipole moments
averaged over the spherically symmetric wave function of the ground
state in the $d$-dimensional space,
%27
\begin{equation}
\psi _{0}(r)=Be^{-r/\sigma },
\end{equation}%
must be, generally speaking, different from zero. The quadrupole moment
equals (see work \cite{R3})
%28
\begin{equation}
\left\langle \hat{Q}_{2}\right\rangle =\left\langle
(3z^{2}-r^{2})\right\rangle =\dfrac{(3-d)(1+d)(d-1)^{2}}{16}.
\end{equation}%
Hence, it is positive (the system is elongated along the $z$-axis)
in the 2$D$ space (as well as in the range $1<d<3$).\,\,At $d>3$, it
is negative (the system is flattened along the $z$-axis) and grows
by absolute value according to the law $d^{4}$.\,\,It is of interest
that the quadrupole and all multipole moments tend to zero if the
space dimensionality tends to 1, which is a consequence of the
collapse.\,\,The next, octupole, moment depending on the
dimensionality $d$ looks like
 %29
\[
\langle\hat Q_4
\rangle=\left\langle\!\frac18(35z^4-30z^2r^2+3r^4)\!\right\rangle=
\]%\vspace*{-7mm}
\begin{equation}
=\frac{3(d-1)^4(1+d)(9-d^2)(5-d)}{2048}.
  \end{equation}
It equals zero in the 3$D$ and 5$D$ spaces and remains positive for
$1<d<3$ and negative for $3<d<5$ (as all even multipole moments do).
At $d>5$, the octupole moment $\left\langle Q_{4}\right\rangle $ is
positive again.\,\,The general formula for nonzero even multipole
moments in the ground state reads
 %30
 \[
 \langle \hat
Q_{2n}\rangle=\frac{(2n-1)!!}{2^{5n}n!}(d-1)^{2n}(d+1)(3^2-d^2)\,\times
\]\vspace*{-7mm}
\begin{equation}
\label{eq:Q2n}
\times\,(5^2-d^2)(7^2-d^2)\mbox{...}\left((2n-1)^2-d^2\right)(2n+1-d).
\end{equation}
A consequence of the general formula (\ref{eq:Q2n}) consists in that
the multipole moments $\left\langle \hat{Q}_{2n}\right\rangle $ in
the spaces with odd dimensionalities equal zero if $d\leq 2n+1$ and
oscillate $n$ times if $d>2n+1$; for even dimensionalities ($d=2n$),
they always differ from zero.\,\,Therefore, all multipole moments
averaged over the spherically symmetric wave functions vanish only
in the 3$D$ space (and, formally, in the 1$D$ space as a result of
the collapse), but it is not so for even dimensionalities and other
odd and fractal ones.\,\,In the 5$D$ space, all multipoles equal
zero in a spherically symmetric field, except for the quadrupole,
which is negative in this case.\,\,In the 7$D$ space, all multipoles
equal zero, but for the quadrupole, which is negative, and the
octupole, which is positive, and so on.\,\,Note also that the
averaging of the multipole expansion (\ref{eq:1/|r-R|}) gives rise
to a factorially divergent asymptotic series, and the character of
divergence grows with the space dimensionality $d$.

Similar regularities are also observed for average multipole moments
in excited states.\,\,For instance, for the first radially excited
state, the Coulomb wave function is
%31
\begin{equation}
\Psi_1(r)=B\left\{\!1-\dfrac{r}{\sigma(\sigma+1)}\!\right\}\exp\left\{\!-\dfrac{r}{\sigma+1}\!\right\}\!.
\end{equation}
The angular part remains the same as for the ground
state.\,\,Therefore, the quadrupole moment is equal to
%32
\begin{equation}
\left\langle 1|\hat{Q}_{2}|1\right\rangle =\frac{(d+1)^{2}(d+11)(3-d)}{16},
\end{equation}
and the octupole moment to
%33
\begin{equation}
\left\langle 1|\hat{Q}_{4}|1\right\rangle
=\frac{3(d+1)^{4}(d+29)(3^{2}-d^{2})(5-d)}{128}
\end{equation}%
with all general regularities being similar to those in the ground
state.\,\,Let us consider a state with the angular moment equal 1
(the $P$ state).\,\,In this case, the wave function of the
one-center problem is
%34
\begin{equation}
\Psi _{P}(r)=Bz\exp \left\{ -\dfrac{r}{\sigma +1}\right\},
\end{equation}
the energy equals $E_{1}=-2/(d+1)^{2},$ and the quadrupole moment
%35
\begin{equation}
\left\langle P|\hat{Q}_{2}|P\right\rangle =\frac{(d+1)^{2}(d+3)(7-d)}{16}
\end{equation}
evidently differs from zero in the 3$D$ geometry. The octupole,
%36
\begin{equation}
\langle
P|\hat{Q}_4|P\rangle=\frac{3(d+1)^4(d+5)(3^2-d^2)(21-d)}{2048},
\end{equation}
and higher multipole moments also oscillate and vanish for the
increasing number of odd $d$-values.\,\,For even $d$-values, all
multipoles are nonzero again.

Making allowance for the dipole interaction $z/R^{2}$ (the second
term in Eq.~(\ref{eq:1/|r-R|})) in the framework of perturbation
theory for the 2$D$ space and at any $d$ (in high-order
terms of the small parameter $1/R^{2}$) deserves a special attention.
In those cases, we obtain a multidimensional Stark problem for a
hydrogen atom in a uniform electric field,
%37
\begin{equation} \label{eq:shtark}
\left\{\! -\frac12\Delta -\frac1r + \epsilon z\!\right\}\Psi = E
\Psi,
\end{equation}
with $\epsilon \equiv 1/R^{2}$.\,\,Let us consider the energy shift
for the ground state, when only the contributions of terms with even
power exponents of $\epsilon $ survive: the Stark effects of the
second, fourth, and so on, orders.\,\,Note that the quadratic Stark
effect was already contained in formula (\ref{eq:u0}) as the fourth
term.\,\,Analogously to the 3$D$ case (see work \cite{R9}), the
hyperparabolic coordinates allow the separation of variables in
Eq.~(\ref{eq:shtark}) to be done, so that, for the angle-independent
ground state, we obtain the system of 1$D$ equations
 %38
\begin{equation} \label{eq:rikkati2}
\begin{array}{l}
\displaystyle \left(\!\xi
\dfrac{d^2}{d\xi^2}+\sigma\dfrac{d}{d\xi}+\dfrac{E}{2}\xi-
\dfrac{\epsilon}{4}\xi^2+\beta_1 \!\right)\chi(\xi)=0,\\[3mm]
\displaystyle \left(\!\eta
\dfrac{d^2}{d\eta^2}+\sigma\dfrac{d}{d\eta}+\dfrac{E}{2}\eta+
\dfrac{\epsilon}{4}\eta^2+\beta_2 \!\right)\phi(\eta)=0,\\[3mm]
\displaystyle \beta_1+\beta_2=1,
\end{array}
\end{equation}
which is identical formally to the equations for the 3$D$ Stark
effect, provided the substitution $1\rightarrow \sigma =$ $=(d-1)/2$
in terms before the first derivative.\,\,After the change of
variables,
%39
\begin{equation} \label{eq:chi_phi}
% \begin{array} {l}
\displaystyle \chi=\exp{\Bigg\{\!\!-\!\int\limits ^\xi\!
F(\xi')d\xi'\!\Bigg\}},\quad \displaystyle
\phi=\exp{\Bigg\{\!\!-\!\int \limits^\eta\! G(\eta')d\eta'\!\Bigg\}}
 %\end{array}
 \end{equation}
the linear second-order equations for the ground state are
reduced to a system of two nonlinear first-order equations of
the Riccati type,
 %40
\begin{equation}
 \begin{array}{l}
\displaystyle
-\dfrac{dF}{d\xi}+F^2-\dfrac{\sigma}{\xi}F+\dfrac{E}{2}-\dfrac{\epsilon}{4}
\xi+\frac{\beta_1}{\xi}=0,\\[3mm]
\displaystyle
-\dfrac{dG}{d\eta}+G^2-\dfrac{\sigma}{\eta}G+\dfrac{E}{2}
+\dfrac{\epsilon}{4}\eta+\frac{\beta_2}{\eta}=0,\\[3mm]
\displaystyle \beta_1+\beta_2=1.
  \end{array}\label{ric1}
\end{equation}
The system of equations (\ref{ric1}) is rather convenient for the
application of perturbation theory (the logarithmic perturbation theory in
the form of recurrent relations), if its solutions are sought in the form of
power series in the small parameter $\epsilon $,
%41
\begin{equation}
\begin{array}{l}
\displaystyle F(\epsilon,\zeta)=\sum\limits_n \epsilon^nF_n(\zeta),\\[5mm]
\displaystyle G(\epsilon,\eta)=\sum\limits_n \epsilon^nG_n(\eta),
\end{array}
\end{equation}
with the coefficient functions taken in the form of
polynomials.\,\,Then, taking only the dipole interaction into
account, we obtain the nonzero contributions to the energy in the
even orders of perturbation theory (the generalized Stark law),
%42
\[
E_0 = -\frac12\sigma^2-\frac1{R^4}\frac{\sigma^4}{8}(1+\sigma)
(5+4\sigma)-\frac1{R^8}\sigma^{10}(1+\sigma)\,\times
\]\vspace*{-7mm}
\begin{equation} \label{eq:E0}
\times\,(192 \sigma^3 +933 \sigma^2 +
1550\sigma+880)+O\left(\!\frac1{R^{12}}\!\right)\!.
\end{equation}
The second term in Eq.~(\ref{eq:E0}) coincides with the fourth one
in formula (\ref{eq:u0}).\,\,Note that series (\ref{eq:E0}) in the
small parameter $1/R^{4}$ also diverges factorially at the fixed
$\sigma =(d-1)/2$.

In view of the results obtained above for multipoles in the
asymptotic expressions for term (\ref{eq:u0}), we note that the
available asymptotics $1/R^{3}$ (the polarizability of the first
order according to perturbation theory) has a quadrupole origin and
is negative (attraction) for $1<d<3$ (in particular, for the 2$D$
problem) and positive (repulsion) for $d>3$.\,\,Moreover, the higher
the space dimensionality, the stronger is the repulsion in the
$1/R^{3}$-asymptotics.\,\,The term of the second order in the
dipole-dipole interaction (the polarizability of the second order)
for the ground state always corresponds to the attraction (the
negative term).

At short distances, where the problem of finding the asymptotics becomes even
more complicated (see work \cite{R2}), the terms have the following
behavior: for the symmetric ground state,
 %43
\begin{equation}
  U_s(R )=-8+64 R^2 \ln\left(\!\frac12 e^{C}R\!\right) +O(R^2),
\end{equation}
where $C=0.57722 ...$ is the Euler constant, for the lowest state of
the antisymmetric term,
%44
\begin{equation}
U_{a}(R)=-\frac{8}{9}-\frac{80}{9}R^{2}+O(R^{4}),
\end{equation}
and, for the first excited symmetric term,
%45
\begin{equation}
U_s(R)=-\frac89+\frac{64}{27}R^2 \ln\left(\!\frac32
e^{C+1}R\!\right) + O(R^2).
\end{equation}

\section{Approximation Formulas for Terms}

For the application of energy terms to be convenient, we will derive
the approximation formulas for the symmetric and antisymmetric
ground terms $V_{s,a}(R)$ in the 2$D$ space.\,\,The approximations
account for the results of direct calculations shown in Figs.~1 and
2, as well as the asymptotic behavior of the terms at short and
large distances discussed in the previous section.

For the symmetric term in the short-distance interval ($R<0.1$), we
use the asymptotics \cite{R1,R2} $V_{s}(R)=1/R-6+64R^{2}\ln
(1/2e^{C}R)$.\,\,At large distances ($R>6.5$), we use asymptotics
(\ref{eq:u0}) and (\ref{eq:deltaE}), when we may put
%46
\[
V_s(R)=-\frac3{32}\frac1{R^3}-\frac{21}{256}\frac1{R^4}-\frac{135}{2048}\frac1{R^5}\,-
\]\vspace*{-5mm}
\begin{equation} \label{eq:V_s(R)}
-\,\frac{159}{1024}\frac1{R^6}-\frac{32}{\pi
e}e^{-2R}\left(\!1+O\left(\!\frac1R\!\right)\!\right)\!.
\end{equation}
In the intermediate region, we use approximation formulas in the
form of the ratio between polynomials of the distance $R$,
 %47
\begin{equation}  \label{eq:V_{appr.}(s)}
V_{s,{\rm appr}}(R)\!=\!\dfrac{b_0 \!+\!b_1R\!+\!b_2R^2
+\!\mbox{...}\!+\!b_{n-2}R^{n-2}
}{R(a_0\!+\!a_1R\!+\!a_2R^2\!+\!\mbox{...}\!+\!a_nR^n)}.
\end{equation}
which should involve two first asymptotic terms at short distances
and all known power-law asymptotics at large ones.\,\,The account of
the power-law asymptotics at large and short distances, as well as
the linearity in the sought parameters in the numerator and the
denominator, imposes the following additional
relations:\vspace*{-2mm}
 %48
\begin{equation} \label{eq:bi=f(a)}
\begin{array}{l}
\displaystyle b_0 =a_0 , \\%[2mm]
\displaystyle b_1 = a_1 -6 a_0 ,\\%[2mm]
\displaystyle b_2 = a_2 -6 a_1 ,\\%[2mm]
\displaystyle b_3 = 1{,}0,\\%[2mm]
\displaystyle b_{n-2} = -(3/32)a_n ,\\%[2mm]
\displaystyle b_{n-3} =-(3/32)a_{n-1}-(21/256)a_{n},\\%[2mm]
\displaystyle b_{n-4} = -(3/32)a_{n-2} \!-\! (21/256)a_{n-1} \!-\!(135/2048)a_n ,\\%[2mm]
\displaystyle b_{n-5} =-(3/32)a_{n-3} - (21/256)a_{n-2}\, -\\%[2mm]
\displaystyle -\,(135/2048)a_{n-1} -(159/1024)a_n.
\end{array}\!\!\!\!\!\!\!\!\!\!\!\!\!\!\!\!\!\!\!\!\!\!
\end{equation}
The parameters $a_{k}$ and $b_{i}$ were determined using the
procedure of best fitting to the calculated term in the range
$0.1<R<6.5$ according to the $\chi _{\mathrm{and}}^{2}$-criterion;
the corresponding $\chi _{\mathrm{and}}^{2}=1.2\times 10^{-6}$. To
obtain a satisfactory approximation in Eq.~(\ref{eq:V_{appr.}(s)}),
it was enough to take $n=9$ and to neglect the last abnormal term in
Eq.~(\ref{eq:V_s(R)}).\,\,In this case, the parameters to fit were
only 10 quantities $a_{k}$ in the denominator of
Eq.~(\ref{eq:V_s(R)}).\,\,Their values are listed in
Table~\ref{tab:approxcoeff}.\,\,All parameters in the numerator of
Eq.~(\ref{eq:V_s(R)}) were unambiguously determined from relations
(\ref{eq:bi=f(a)}).

%Табл. 2
\begin{table}[b]
 \noindent\caption{Approximation coefficients\\ for
symmetric and antisymmetric terms}\vskip3mm\tabcolsep13.2pt

\noindent{\footnotesize\begin{tabular}{|c|c|c|c|}
 \hline \multicolumn{1}{|c|}
{\rule{0pt}{5mm}$i$ } & \multicolumn{1}{|c|}{$a_i,s$}&
\multicolumn{1}{|c|}{$a_i,a$}&
\multicolumn{1}{|c|}{$b_i,a$}\\[2mm]%
\hline%
\rule{0pt}{5mm}0& 0.13241& 1.5067& \\
1& 1.84394& 0.9674& \\
2& 4.12047& 1.3044& \\
3& 1.30497& 0.073~\,& \\
4& 1.29747& 2.5425& $-2{.}1696$~~\, \\
5& 0.03477& 0.8491&0.4898\\
6& $-0{.}00581$~~\,& $-0{.}3083$~~\,& \\
7& 0.19956& $-3{.}4974$~~\,& \\
8& $-0{.}06119$~~\,  & 3.0725& 1.8518 \\
9& 0.01156   & ~~\,\,0.072605&\\[2mm]
 \hline
\end{tabular}}
\label{tab:approxcoeff}
\end{table}

An approximation formula similar to Eq.~(\ref{eq:V_{appr.}(s)}) was
used for the antisymmetric term:\vspace*{-2mm}
%49
\[
V_{a,{\rm appr}}(R)=\{b_0 +b_1R+b_2R^2 +\mbox{...}+b_{n-2}R^{n-2}+
\]\vspace*{-8mm}
\[
+b_{n-1} \;32 R^{n+1}\exp(-2R-1)/\pi\}/
\]\vspace*{-8mm}
\begin{equation}  \label{eq:V_{appr.}(a)}
/\{R(a_0+a_1R+a_2R^2+\ldots+a_nR^n)\}.
\end{equation}

%Fig.4
\begin{figure}[t]
\vskip1mm
\includegraphics[width=\column]{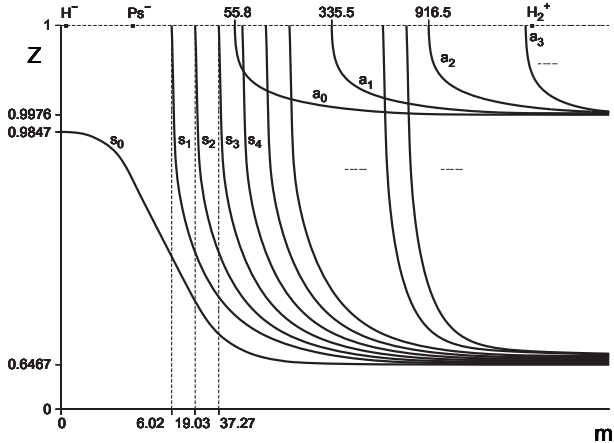} %snew.eps
\vskip-3mm\caption{Diagrams (schematically) of the energy level
stability for a two-dimensional system of three charged particles}
\label{fig:stabilitydiagrams}
\end{figure}

%Табл. 3
\begin{table}[h!]
\vskip5mm \noindent\caption{Critical mass values\\
\boldmath$m_{\mathrm{crit}.}^{(n)}$ for symmetric states with (the
second row) and without (the third row) regard for the abnormal
centripetal attraction $-\frac{1}{4r^{2}}$}\vskip3mm\tabcolsep3.7pt

\noindent{\footnotesize\begin{tabular}{|l|c|c|c|c|c|c|}
 \hline \multicolumn{1}{|c|}
{\rule{0pt}{5mm}State $ n $} &
\multicolumn{1}{|c|}{0}&\multicolumn{1}{|c|}{1}&
\multicolumn{1}{|c|}{2}&\multicolumn{1}{|c|}{3}&\multicolumn{1}{|c|}{4}&\multicolumn{1}{|c|}{5}\\[2mm]%
\hline%
\rule{0pt}{5mm}Total potential& 0 & 6.01 & 19.03 & 37.27 & 59.03 & 84.43\\%
without $-(1/4)/r^2$&0.75 & 8.09 & 22.19 & 42.67 & 68.92 & 99.89 \\%
3-part. calc.&0&5.9~\,&20.2~\, & 40.9~\, &67.2~\, &98.3~\,\\[2mm]%
\hline
\end{tabular}}\label{tab:mscrit}\vspace*{-4mm}
\end{table}%
%Табл. 4
\begin{table}[h!]
\vskip5mm \noindent\caption{Critical mass values
\boldmath$m_{\mathrm{crit}.}^{(n)}$\\ for antisymmetric states with
(the second row)\\ and without (the third row) regard for the
abnormal\\ centripetal attraction
$-\frac{1}{4r^{2}}$}\vskip3mm\tabcolsep4.7pt

\noindent{\footnotesize\begin{tabular}{|l|c|c|c|c|c|}
 \hline \multicolumn{1}{|c|}
{\rule{0pt}{5mm}State $ n $} &
\multicolumn{1}{|c|}{0}&\multicolumn{1}{|c|}{1}&
\multicolumn{1}{|c|}{2}& \multicolumn{1}{|c|}{3}&
\multicolumn{1}{|c|}{4}\\[2mm]%
\hline%
\rule{0pt}{5mm}Total potential& ~\,55.8  & 335.5 & ~\,916.5 & 1741.5 & 2807 \\
without $-(1/4)/r^2$ & 128.3  & 509.2 & 1155.5 & 2063.5 & 3229 \\
3-part. calc. & 111~~  & 643~~ & 1675~~ & 3266~~ & 5690 \\[2mm]
\hline
\end{tabular}}\label{tab:macrit}\vspace*{-4mm}
\end{table}%

\noindent The difference consists in that the abnormal asymptotics
(the last term with the opposite sign) from Eq.~(\ref{eq:V_s(R)}) is
taken now into account, and we have the short-distance asymptotics
$V_{a}(R)=1/R+10/9+O(R^{2})$.\,\,This means that, in the second and
third formulas of Eqs.~(\ref{eq:bi=f(a)}), the coefficient $-6$
should be better substituted by $10/9$.\,\,In addition, we neglect
the last two relations from Eqs.~(\ref{eq:bi=f(a)}) in the numerator
of Eq.~(\ref{eq:V_{appr.}(a)}).\,\,The other parameters in the
numerator of Eq.~(\ref{eq:V_{appr.}(a)}) are unambiguously
determined by relations (\ref{eq:bi=f(a)}).\,\,The obtained values
of approximation parameters (at $n=9$) for the antisymmetric term
are quoted in Table~\ref{tab:approxcoeff} (the third and fourth
columns).\,\,We hope for that the accuracy of the approximations
obtained for both symmetric and antisymmetric terms is high enough
for those terms to be used in further researches.

It is worth noting that the symmetric term in the 2$D$ problem
decreases according to the asymptotic law $-1/R^{3}$ in the
preasymptotic region of large distances.\,\,As a result, there is no
reason for the realization of the preasymptotic law $-1/R^{2}$,
which was discussed in work \cite{R10} for neutral atoms in the
3$D$~case.%\vspace*{-2mm}

\section{Stability Diagrams\\ in the Born-Oppenheimer\\ Adiabatic Approximation}

The determined terms allow us, in accordance with
Eq.~(\ref{eq:eq9}), to find vibration spectra in the BO
approximation and to plot the corresponding stability diagrams (for
the three-particle problem in the 2$D$ space, they were found in
work \cite{R3}).\,\,In Fig.~\ref{fig:stabilitydiagrams}, the curves
of stability threshold calculated for the symmetric $(s_{n})$ and
antisymmetric $(a_{n})$ states in the BO adiabatic approximation are
exhibited schematically.\,\,The corresponding curves on the plane
$(m,Z)$ mean that the $n$-th symmetric state exists to the right
from the corresponding curve $s_{n}$, whereas the $n$-th
antisymmetric state exists to the right from the corresponding curve
$a_{n}$.\,\,Certainly, although the diagrams are plotted for all
values of mass ratio $m$, one should bear in mind that the BO
adiabatic approximation is physically justified only for the
molecular mode, so that the results obtained for small $m$-values
are shown only to make a more complete comparison with the results
of direct three-particle calculations taken from work \cite{R3}.
Tables~\ref{tab:mscrit} and \ref{tab:macrit} also demonstrate the
results of calculation, according to simple Eq.~(\ref{eq:eq9}), of
critical masses $m_{\mathrm{crit}.}^{(n)}$ for the symmetric and
antisymmetric states, when $Z=1$ and the energy $\varepsilon =0 $.
Those values are partially shown in
Fig.~\ref{fig:stabilitydiagrams}: they determine vertical
asymptotes.\,\,We also calculated the simple 1$D$ problem
(\ref{eq:eq9}) with the effective potentials
$V_{\mathrm{eff}}(r)=V(r)-(1/4/(m+1/2))/r^{2}$ both in the full
variant (the second rows in Tables~\ref{tab:mscrit} and
\ref{tab:macrit}) and neglecting the second term associated with the
kinetic energy of the centripetal attraction (Eq.~(\ref{eq:eq9}))
(the third rows in Tables~\ref{tab:mscrit} and \ref{tab:macrit}).
The last rows show the results of three-particle calculations for
the critical values $m_{\mathrm{crit}.}^{(n)}$ (more accurate in
comparison with work \cite{R3} and carried out using a Gaussian
basis with 1300 components).

\vspace*{-1.0mm}Let us compare various variants of critical mass,
$m_{\mathrm{crit}.}$, sets in more details.\,\,First of all, we
should pay attention to that the BO adiabatic approximation is a
variational estimate \textquotedblleft from above\textquotedblright\
for the energies and critical values $m_{\mathrm{crit}.}^{(n)}$.
Therefore, all the curves in Fig.~\ref{fig:stabilitydiagrams}
calculated in the BO approximation must lie above those calculated
in the three-particle problem and be more correct if the mass $m$ is
larger.\,\,From a comparison of the second and fourth rows in
Table~\ref{tab:mscrit}, it follows that, while calculating the
symmetric state in the framework of the three-particle scheme, the
accuracy higher than that attained in the BO approximation was
obtained only for the first two states.\,\,Hence, the calculation of
the terms and the vibration level energies in the adiabatic
approximation turns out justified from the viewpoint of accuracy and
the understanding of regularities.\,\,To a larger extent, those
remarks concern anomalously weakly coupled antisymmetric states.
From a comparison of the second and fourth rows in
Table~\ref{tab:mscrit}, it follows that even the substantially
corrected, in comparison with work \cite{R3}, results of
calculations (using the basis of about 1100 Gaussoid-like
components) turn out unsatisfactory at the quantitative level in
comparison with the results obtained in the BO approximation.\,\,At
last, a comparison of the results in the second (the total effective
term) and third (here, the centripetal attraction $-(1/4)/r^{2}$ is
neglected) rows of Tables~\ref{tab:mscrit} and \ref{tab:macrit}
allows us to evaluate the order of the contribution given by this
centripetal attraction to the stability diagram structure.\,\,For
instance, neglecting the centripetal attraction in the symmetric
ground state always gives rise to a finite critical mass, whereas,
for higher excitations, the difference between the results in the
second and third rows becomes larger and larger.\,\,A special
attention should be paid to the fact that, in the 2$D$ problem and
in the adiabatic approximation for the ground state, we obtain a
bound three-particle level at any mass value, which follows from the
known fact \cite{R12,R13} that, in the case of two particles and an
attractive potential, there always exists at least one weakly bound
state with the energy that exponentially depends on the potential,
$E_{0}=-(\hbar ^{2}/mr_{0}^{2})\exp (\mathrm{Const}/v_{0}+C2)$,
where $v_{0}$ is the zero-momentum Fourier component of the
attractive potential, $v_{0}<0$.

Concerning the horizontal asymptotes in Fig.~\ref{fig:stabilitydiagrams},
the threshold curves determine the minimum value of charge
%50
\begin{equation}
Z_{\mathrm{crit}} =\min\left\{[1-RV_{s}(R)]^{-1}\right\},
\end{equation}
for symmetric and antisymmetric states in the limit of large mass
$m$.\,\,For the corresponding $R_{\mathrm{crit}(s)}=0.89$ in the
symmetric state, we obtain $Z_{\mathrm{crit}(s)}=0.64686$, and this
value coincides with the results of three-particle calculations
\cite{R3}.\,\,Accordingly, the antisymmetric threshold curves have
the minimum $Z_{\mathrm{crit}(a)}=$ $=0.9976$, and
$R_{\mathrm{crit}(a)}=5.7$.\,\,The indicated values were determined
more exactly than it could be done in three-particle calculations.

Note also that the results of three-particle calculations according
to Eq.~(\ref{eq:linear_eqs}) imply that there are no antisymmetric
bound states for all masses at dimensionalities $d>3.337$.\,\,This
is a result of the repulsion provided by both the centrifugal
barrier $l_{\mathrm{eff}}(l_{\mathrm{eff}}+1)/R^{2}$, where
$l_{\mathrm{eff}}=(d-3)/2$ is the effective angular hypermomentum,
and the repulsive asymptotics $\left\langle Q_{2}\right\rangle
/2R^{3}$ of term (\ref{eq:u0}) with the abnormal quadrupole moment
of a hydrogen atom.\,\,In this case, as the dimensionality grows,
the antisymmetric term becomes exclusively repulsive (it remains
attractive only in the 2$D$ and 3$D$ problems, if the space
dimensionality is an integer number).\,\,In turn, the results of
three-particle calculations according to Eq.~(\ref{eq:linear_eqs})
demonstrate that symmetric states are absent for all masses $m$ only
at large space dimensionalities, when $d>9$, and the bound state of
an atomic hydrogen ion is absent at $d>6$.\,\,This follows  from the
presence of the centrifugal barrier $(d-3)(d-1)/4R^{2}$, abnormal
quadrupole moment, and repulsion $\left\langle Q_{2}\right\rangle
/2R^{3}$.

Note at last that the results of calculations for the critical
masses $m_{\mathrm{crit}(s)}$ of various states in the BO
approximation can be approximated by a square law, depending on the
number of a state (in the case of 3$D$ problem, this was done in
work \cite{R3}).\,\,In particular, for the symmetric states,
%51
\begin{equation}
m_{\mathrm{crit}(s),n}=2.47(n+1)^{2}-3.3.  \label{eq:2.89}
\end{equation}%
This formula is much more exact than the approximation obtained from
three-particle calculations \cite{R3}, especially for high excited states.
The general constant at $n^{2}$ in Eq.~(\ref{eq:2.89}) is determined as the
asymptotics of the quasiclassical approximation
%52
\begin{equation}
m_{\mathrm{crit}}=(\pi /J)^{2}n(n+1),  \label{eq:mcrit}
\end{equation}%
in which the quasiclassical integral (neglecting the centripetal
attraction)\vspace*{-2mm}
\[
J=1.84785
\]\vspace*{-5mm}

\noindent and, accordingly,\vspace*{-2mm}
\[
\left( \!\frac{\pi }{J}\!\right) ^{\!2}=2.89
\]
\vspace*{-5mm}

\noindent are determined only by the negative part of the term.
Similarly, for the antisymmetric critical states, we obtain the
approximation\vspace*{-2mm}
%53
\begin{equation}
m_{\mathrm{crit}(s),n}=138.6n(n+1)+51, \label{eq:macrit}
\end{equation}%
\vspace*{-5mm}

\noindent and, from the quasiclassical approximation for
antisymmetric states,\vspace*{-2mm}
\[
\left(\! \frac{\pi }{J}\!\right) ^{\!2}=129.845.
\]
\vspace*{-5mm}

\noindent Note that the quasiclassical estimation for the
asymptotics of energy levels in the 2$D$ case is not
well-grounded.\,\,There are principal difficulties in the estimation
of quasiclassical integrals at both large and short distances owing
to the centripetal attraction $-1/(4R^{2})$.

As a consequence of those approximations, it follows from the
adiabatic approximation that there are 26 excited levels for a
molecular hydrogen ion H$_{2}^{+}$ (with the mass $m=1836.152701)$
in the symmetric state.\,\,In the antisymmetric state, the total
number of levels equals four.%\vspace*{-2mm}

\section{Final Remarks}

To summarize, we would like to note that the researches carried out
for three charged particles in the framework of Born--Oppenheimer
approximation allowed us to establish a number of abnormal
regularities arising in the 2$D$ space and the spaces of arbitrary
dimensionality.\,\,In the 2$D$ space, the multipole expansions for
Coulomb potentials in a spherically symmetric field are nonzero.
Also nonzero are the quadrupole, octupole, and other multipole
moments in the $d$-dimensional problems.\,\,In the 2$D$ problem, the
quadrupole moment of a hydrogen atom is positive and generates an
attractive asymptotics $\sim$$ -1/R^{3}$ for the ground-state term,
whereas, in the 3$D$ problem, this contribution equals zero.\,\,As a
result, it was found that a hydrogen atom in the 2$D$ space is
polarizable already in the first-order perturbation theory.
Expressions for higher multipole moments in spaces with arbitrary
dimensionalities are obtained and analyzed.\,\,The antisymmetric
terms of trions $XXY$ are attractive only in the 2$D$ and 3$D$
problems.\,\,Moreover, it is shown that there is a critical
dimensionality value $d_{\mathrm{crit}}=3.337$, and there are no
bound antisymmetric states in spaces with
$d>d_{\mathrm{crit}}$.\,\,The abnormal behavior of the asymptotics
for interaction potentials of the van der Waals type between neutral
hydrogen atoms in the 2$D$ space is demonstrated.

The convenient approximation formulas for 2$D$ terms are
proposed.\,\,In the framework of the Born--Oppenheimer adiabatic
approximation, the stability diagrams for the 2$D$ space are
obtained, and the main characteristic asymptotics for the stability
threshold curves for trions $XXY$ are plotted and analyzed.\,\,They
agree with the stability diagrams obtained earlier in three-particle
calculations.

\vskip2mm

\textit{The authors express their gratitude to
\fbox{M.V.\,Kuzmenko\!}, who participated in this research
at\linebreak the initial stage, and to B.E.\,Grynyuk for the
use\-ful discussion of effective numerical cal\-cu\-la\-ti\-on
proce-\linebreak dures.}

\rezume{%
І.В.\,Сименог, В.В.\,Михнюк,
Ю.М.\,Бідасюк\vspace*{-1mm}}{ЕНЕРГЕТИЧНІ ТЕРМИ ТА ДІАГРАМИ
\\СТАБІЛЬНОСТІ  ДЛЯ  2$D$ ЗАДАЧІ \\ТРЬОХ ЗАРЯДЖЕНИХ ЧАСТИНОК} {Для
двовимірних кулонівських систем типу симетричних тріонів $XXY$ у
варіаційному підході отримано симетричний та антисиметричний терми.
Дано якісне пояснення діаграм стабільності та певних аномалій в $2D$
просторі на основі адіабатичного наближення Борна--Опенгаймера.
Виконано аналіз отриманих для довільної вимірності простору
асимптотик енергетичних термів на великих відстанях, і запропоновано
апроксимаційні формули для $2D$ термів. Встановлено аномальну
залежність мультипольних моментів від вимірності простору у випадку
сферично-симетричного поля. Проведено кількісне порівняння  основних
результатів для  $2D$ і $3D$ задач двох кулонівських центрів.}

%TCIMACRO{%
%\TeXButton{Figs and Tabs}{\begin{figure}\caption{}\label
%{fig:sterm}\end{figure}
%\begin{figure}\caption{}\label{fig:aterm}\end{figure}
%\begin{figure}\caption{}\label{fig:saterms}\end{figure}
%\begin{figure}\caption{}\label{fig:stabilitydiagrams}\end{figure}
%\begin{figure}\caption{}\label{Fig5}\end{figure}
%\begin{figure}\caption{}\label{Fig6}\end{figure}
%\begin{figure}\caption{}\label{Fig7}\end{figure}
%\begin{figure}\caption{}\label{Fig8}\end{figure}
%\begin{figure}\caption{}\label{Fig9}\end{figure}
%\begin{table}\caption{}\label{tab:saparameters}\end{table}
%\begin{table}\caption{}\label{tab:approxcoeff}\end{table}
%\begin{table}\caption{}\label{tab:mscrit}\end{table}
%\begin{table}\caption{}\label{tab:macrit}\end{table}
%\begin{table}\caption{}\label{tab5}\end{table}
%\begin{table}\caption{}\label{tab6}\end{table}
%\begin{table}\caption{}\label{tab7}\end{table}
%\begin{table}\caption{}\label{tab8}\end{table}
%\begin{table}\caption{}\label{tab9}\end{table}}}%
%BeginExpansion
%\begin{figure}\caption{}\label{fig:sterm}\end{figure}
%\begin{figure}\caption{}\label{fig:aterm}\end{figure}
%\begin{figure}\caption{}\label{fig:saterms}\end{figure}
%\begin{figure}\caption{}\label{fig:stabilitydiagrams}\end{figure}
%\begin{table}\caption{}\label{tab:saparameters}\end{table}
%\begin{table}\caption{}\label{tab:approxcoeff}\end{table}
%\begin{table}\caption{}\label{tab:mscrit}\end{table}
%\begin{table}\caption{}\label{tab:macrit}\end{table}
%EndExpansion

\end{document}